\documentclass[preprint, showpacs, reprintnumbers, amsmath, amssymb, superscriptaddress]{revtex4-2}
\usepackage{epsf}
\usepackage{graphicx}
\usepackage{sidecap}

 \usepackage{soul}
 \usepackage{array}
 \usepackage{amsmath}
 \usepackage{amssymb}
 \usepackage{dcolumn}
 \usepackage{epstopdf}
 \usepackage{bm}
 \usepackage{gensymb}
 \usepackage{amsmath}

\usepackage{color}
\usepackage{hyperref}
\usepackage{soul}
\sethlcolor{green}
\hypersetup{
    colorlinks=true,
    citecolor=red,
    linkcolor=red,
    filecolor=blue,   
    urlcolor=blue,
}
 
\def \beq {\begin{equation}}
\def \eeq {\end{equation}}
\begin{document}

\onecolumngrid

\begin{center}
 
  \textbf{\Large Electronic structure in a transition metal dipnictide $\mathrm{\textbf{TaAs}_2}$}\\[.2cm]
Sabin~Regmi$^{1,2,*}$, Cheng-Yi Huang$^3$, Mojammel A. Khan$^{4}$, Baokai Wang$^3$, Anup Pradhan Sakhya$^1$, M. Mofazzel Hosen$^1$, Jesse Thompson$^1$, Bahadur Singh$^5$, Jonathan D. Denlinger$^6$, Masahiro Ishigami$^1$, J.F. Mitchell$^4$, Dariusz Kaczorowski$^7$, Arun Bansil$^3$, and Madhab Neupane$^{1,\dagger}$\\[.2cm]
 {\itshape
    $^{1}$Department of Physics, University of Central Florida, Orlando, Florida  32816, USA\\
    $^2$Currently at Idaho National Laboratory, Idaho Falls, Idaho 83415, USA\\
    $^3$Department of Physics, Northeastern University, Boston, Massachusetts 02115, USA\\
  	$^4$Materials Science Division, Argonne National Laboratory, Lemont, Illinois 60439, USA \\
  	$^5$Department of Condensed Matter Physics and Materials Science, Tata Institute of Fundamental Research, Mumbai 400005, India\\
  	$^6$Lawrence Berkeley National Laboratory, Berkeley, California 94720, USA\\
  	$^7$Institute of Low Temperature and Structure Research, Polish Academy of Sciences, Okólna 2, 50-422 Wrocław, Poland
  }
\\[.2cm]
$^*$Corresponding author: sabin.regmi@inl.gov\\
$^{\dagger}$Corresponding author: madhab.neupane@ucf.edu
\\[1cm]
\end{center}
 
\begin{abstract}
\textbf{The family of transition-metal dipnictides (TMDs) has been of theoretical and experimental interest because this family hosts topological states and extremely large magnetoresistance (MR). Recently, $\mathrm{TaAs}_2$, a member of this family, has been predicted to support a topological crystalline insulating state.  Here, by using high-resolution angle-resolved photoemission spectroscopy (ARPES), we reveal both closed and open  pockets in the metallic Fermi surface and linearly dispersive bands on the ($\overline{2}01$) surface, along with the presence of extreme MR observed from magneto-transport measurements. A comparison of the ARPES results with first-principles computations shows that the linearly dispersive bands on the measured surface of $\mathrm{TaAs}_2$ are trivial bulk bands. The absence of symmetry-protected surface state on the  ($\overline{2}01$) surface indicates its topologically dark nature. The presence of open Fermi surface features suggests that the open-orbit fermiology could contribute to the extremely large MR of $\mathrm{TaAs}_2$.}
\end{abstract}
\maketitle

Nontrivial insulators with spin-polarized surface states - the topological insulators (TIs) \cite{HasanKane2010, QiZhang2011, HasanXuNeupane2015, BansilLinDas2016} - spurred an intense interest in  topological quantum materials in view of their novel physics and potential applications. In three dimensions, the $\mathbb{Z}_2$ topological invariants ($\nu_0$:$\nu_1\nu_2\nu_3$) serve as quantum numbers classifying the strong and weak nontriviality of the TIs \cite{FuKane2007, FuKaneMele2007}. The non-zero value of the strong topological index ($\nu_0$ = 1) distinguishes strong TIs from other insulators, while the weak TIs are characterized by non-zero value of at least one of the weak topological indices ($\nu_1\nu_2\nu_3$). The gapless surface states in strong TIs form an odd number of Dirac cones and are robust as long as the time-reversal symmetry is preserved \cite{HasanKane2010, QiZhang2011, HasanXuNeupane2015, Xia2009, Zhang2009, Chen2009}. In contrast, weak TIs feature an even number of Dirac cones on specific surfaces which are not robust against disorder, with the remaining surfaces being topologically dark, i.e., without topological surface state \cite{FuKane2007, FuKaneMele2007, Noguchi2019}. In the so-called topological crystalline insulators (TCIs), the crystalline symmetries such as the mirror or rotational symmetry can protect the Dirac nodes \cite{Fu2011, Hsieh2012, Tanaka2012, Xu2012}. In such TIs, the Dirac nodes may appear at  generic momentum positions away from the time-reversal invariant momentum (TRIM) points \cite{Hsieh2012, Tanaka2012, Xu2012, Zhou2018, Hsu2019, Wang2019}.\\

Transition metal dipnictides (TMDs) have been attracting research interest due to their potential to host topological states \cite{Wang2019, Xu2016, Luo2016, Autes2016, Gresch2017, Chen2017, Shao2019, Sims2020, Bannies2021} and exhibit extremely large magnetoresistance (MR) \cite{Bannies2021, Dhakal2021, Wang2014, Wu2016, Yuan2016, Shen2016,  Wang2016, Li2016, Kumar2017, Lou2017, Yokoi2018, Singha2018, Butcher2019, Chen2021}. TMDs possess multiple crystalline symmetries, such as the mirror and rotational symmetries that can support various crystalline symmetry protected topological states in their electronic structures \cite{Wang2019, Xu2016}. $TnPn_2 ~~ (Tn = \mathrm{Nb, Ta}; ~~ Pn = \mathrm{P, As, Sb})$, which belong to the TMD family, have been predicted to feature weakly non-trivial electronic structure \cite{Wang2019, Xu2016, Luo2016} characterized by non-trivial weak topological indices ($\nu_1\nu_2\nu_3$) = (111).  The extremely large values of MR in quantum materials have been reported to originate from various factors, including the compensation of electron-hole density \cite{Chen2017, Ali2014, Zeng2016, Hosen2020}, open Fermi surface (FS) orbits \cite{Lou2017, Wu2020}, and  ultrahigh mobility \cite{ Mun2012, Tafti2016}. Within the TMD family, there is no consensus on the origin of the extremely large MR with suggestions ranging from   electron-hole compensation  \cite{Chen2017, Yuan2016, Wu2016, Wang2017} to open-orbits in the FS \cite{Lou2017} and  other possible factors such as high residual resistivity ratio and ultrahigh mobility \cite{Wang2014, Kumar2017, Matin2018}.\\

$\mathrm{TaAs}_2$, which is a member of the TMD family, respects both time-reversal and inversion symmetries, and it has been predicted to be a weak TI \cite{Wang2019, Xu2016, Luo2016} and a rotational symmetry protected TCI with type-II Dirac dispersion on the (010) surface \cite{Wang2019}, along with a large unsaturating MR \cite{Luo2016}. Therefore, a comprehensive investigation to understand the topological nature of its electronic structure is desirable. A previous theoretical study predicts equal volume of electron and hole pockets in the FS indicative of a near electron-hole compensation in this material \cite{Wang2019}. Some magnetotransport  reports also suggest $\mathrm{TaAs}_2$ to have almost complete electron-hole compensation with high carrier mobility \cite{Wu2016, Yuan2016}, while others report a finite imbalance between the electron and hole densities suggesting $\mathrm{TaAs}_2$ as a not perfectly compensated semimetal \cite{Luo2016}.  The contribution of other factors to the extremely large MR in $\mathrm{TaAs}_2$ thus warrants further investigation. \\

Here, by using high-resolution angle-resolved photoemission spectroscopy (ARPES)  in conjunction with density functional theory (DFT) computations, we report the detailed study of the electronic structure of $\mathrm{TaAs}_2$.  Our study elucidates the metallic FS of $\mathrm{TaAs}_2$, which comprises of closed electron and hole pockets as well as some non-closing features. Linearly dispersing states are observed  on the cleaved ($\overline{2}01$) surface, which is different from the (010) surface predicted to feature rotational-symmetry protected surface state. A careful comparison of the experimental and computational results confirms that the observed linear bands are, in fact, trivial bulk bands.  Our magneto-transport measurements reveal extremely large MR. The presence of non-closing feature in the FS could point to the possibility of open-orbit fermiology being a contributing factor towards such large MR. Our study provides a framework for a deeper understanding of the electronic structure and the origin of the unusual MR in the TMD-family.\\

Stoichiometric amounts of elemental Ta and high-purity As were put in two different alumina crucibles and were subsequently placed in a fused silica tube. High purity $\mathrm{I}_2$ was used as a growth agent with a typical concentration of 2 mg/cm$^3$. The tube was then sealed under vacuum and the crystals were grown in a temperature gradient of 900\degree C (Source) to 850\degree C (Sink) for 7 days. Large plate like crystals with length of 3-7 mm were obtained with some redundant arsenic grains in the ampule.  The crystal structure was determined by X-ray diffraction on a Kuma-Diffraction KM4 four-circle diffractometer equipped with a CCD camera using Mo K$\alpha$ radiation, while chemical composition was checked by energy dispersive X-ray analysis performed using a FEI scanning electron microscope equipped with an EDAX Genesis XM4 spectrometer. Transport measurements were performed in a Quantum Design Physical Property Measurement Systems (PPMS). Gold wire (25 micron) contacts were placed on the sample using Epotek H20E Epoxy. A four-probe contact method was used for the AC resistivity measurement with an excitation current of 3 mA at a frequency of 57.9 Hz. Electrical resistivity measurements were performed in a temperature range of 2-250 K and MR measurements were performed at 1.8 K and an applied field up to 9 T.\\

Synchrotron-based ARPES measurements of the electronic structure were performed at the Advanced light Source beamline 10.0.1.1 at the Berkeley national laboratory, which is equipped with a Scienta R4000 hemispherical electron analyzer. Single crystals were cleaved \textit{in-situ} in an ultra-high vacuum condition (better than 10$^{-10}$ Torr). The cleaved crystal surfaces did not degrade within the time period of our measurements. The energy resolution was set to be better than 20 meV and the angular resolution was set to be better than 0.2\degree for the synchrotron measurements.  Measurements were conducted at a temperature of 18 K. \\

   \begin{figure*}
	\centering
	\includegraphics[width=0.99\textwidth]{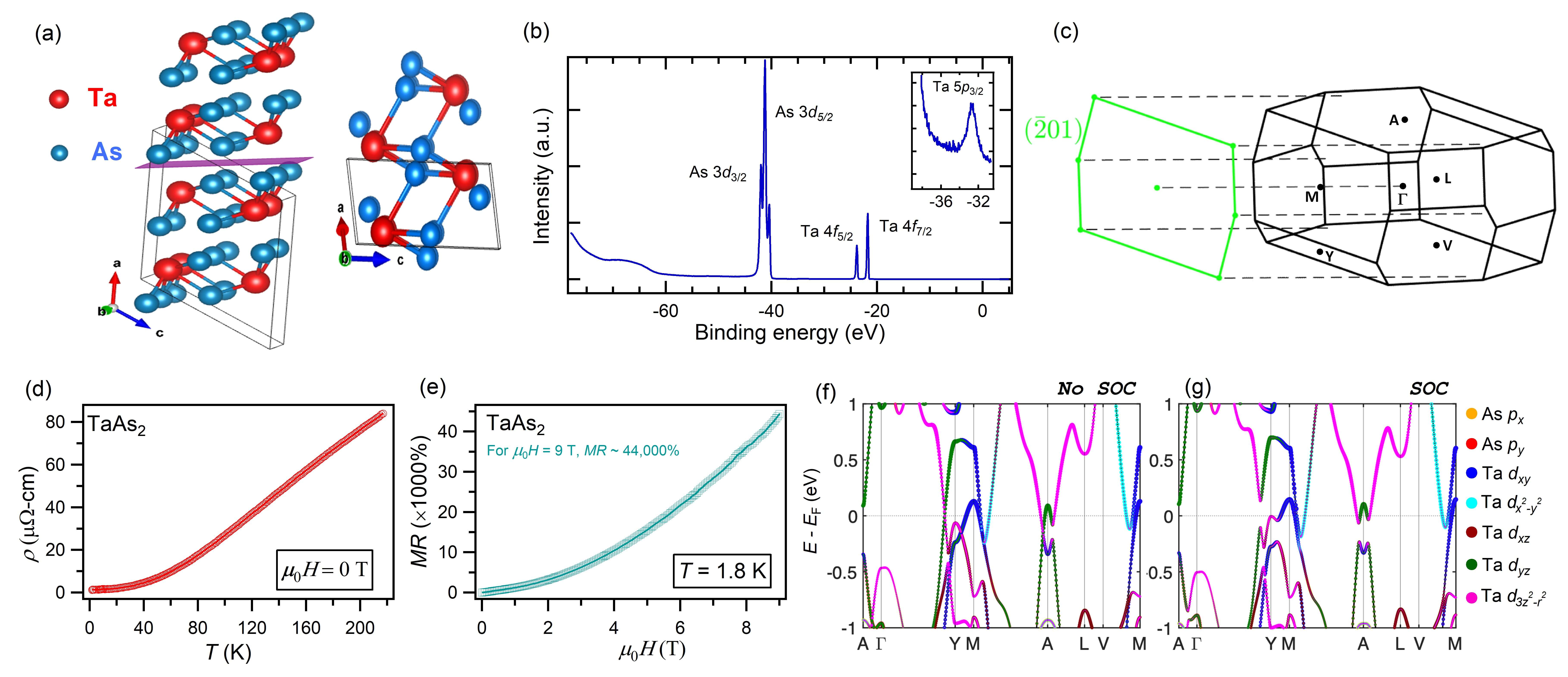}
	\caption{{Crystal structure, bulk characterization, and the computed band structures.} (a) Monoclinic crystal structure of $\mathrm{TaAs}_2$ with conventional unit cell (left) and primitive unit cell (right). Red and blue balls represent Ta and As atoms, respectively.  Purple plane represents the ($\overline{2}01$) plane. (b) Spectroscopic core level spectrum. The inset shows the zoomed-in view around the region pointed to by the arrow. (c) Primitive bulk BZ labeled with the high symmetry points. A two-dimensional BZ projected onto the ($\overline{2}01$) surface is also shown.  (d) Temperature dependence of the resistivity in the absence of applied field. (e) Unsaturated  MR up to 9 T at a temperature of 1.8 K. (f)-(g) Bulk band calculations along the high symmetry directions without and with spin-orbit coupling.}
	\label{Fig1}
\end{figure*}

First-principles calculations based on density functional theory (DFT) \cite{HK64, KS65} were performed using the projector augmented wave (PAW) method \cite{Blochl94} as implemented in the  Vienna ab initio Simulation Package ({\sc VASP}) \cite{Kresse1,Kresse2,Kresse3, Kresse4}. Exchange-correlation effects were treated using the generalized gradient approximation  (GGA)  \cite{PBE}. An energy cutoff of 350 eV was used for the plane-wave basis set and a $\Gamma$-centered 12$\times$12$\times$8 k-mesh was used for  Brillouin zone(BZ) integrations. The atomic positions were relaxed until the residual forces were less than 0.001 eV/$\AA$ on each atom. A tight binding model with atom centered Wannier functions was constructed using the {\sc VASP2WANNIER90} interface \cite{Mostofi2008}. The constant energy contours and the energy-momentum spectrum projected on to the ($\overline{2}01$) surface were obtained using  the iterative Green's function method, employing the WannierTools package \cite{Wu2018}. \\

\begin{figure*}
\centering
\includegraphics[width=1\textwidth]{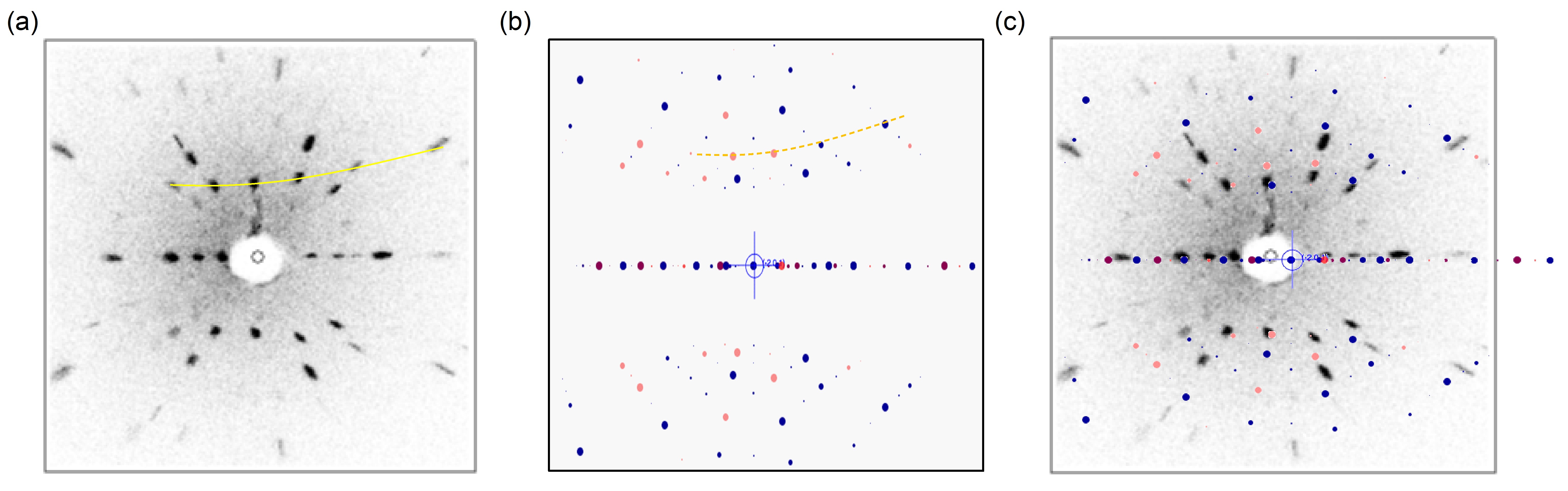}
\caption{ {Laue diffraction from $\mathrm{Ta}\mathrm{As}_2$. (a) The Laue diffraction. (b) QLaue simulation. (c) Comparison of experimental result with QLaue simulation.}}
\label{Fig2}
\end{figure*}
 
$\mathrm{TaAs}_2$ crystallizes in a monoclinic structure (Fig. 1(a)) with space group number 12 (\textit{C2/m}). With a non-magnetic ground state, $\mathrm{TaAs}_2$ respects the mirror symmetry, a two fold (\textit{C}$_{2[010]}$) rotational symmetry, and the inversion symmetry \cite{Wang2019, Xu2016, Wu2016, Butcher2019}. Figure \ref{Fig1}(b) shows the spectroscopic core level spectrum of $\mathrm{TaAs}_2$. Sharp peaks of As 3\textit{d}$_{3/2}$, As 3\textit{d}$_{5/2}$, Ta 4\textit{f}$_{5/2}$ and Ta 4\textit{f}$_{7/2}$ are observed along with a small peak of Ta 5\textit{p}$_{3/2}$ (see inset). Figure \ref{Fig1}(c) shows the bulk primitive BZ with high symmetry points marked by blue dots. Temperature-dependent resistivity plot shows the semimetallic/metallic nature down to 1.8 K (Fig. \ref{Fig1}(d)). Figure \ref{Fig1}(e) shows MR as a function of  the applied field. The MR  remains unsaturated up to 9 T, and at a temperature of 1.8 K with an applied field of 9 T, a colossal MR $\approx$ 44,000$\%$ is observed, which is compatible with the value reported in semimetallic materials such as $\mathrm{WTe}_2$ \cite{YWang2016}. \\

\begin{figure*}
 	\centering
 	\includegraphics[width=1\textwidth]{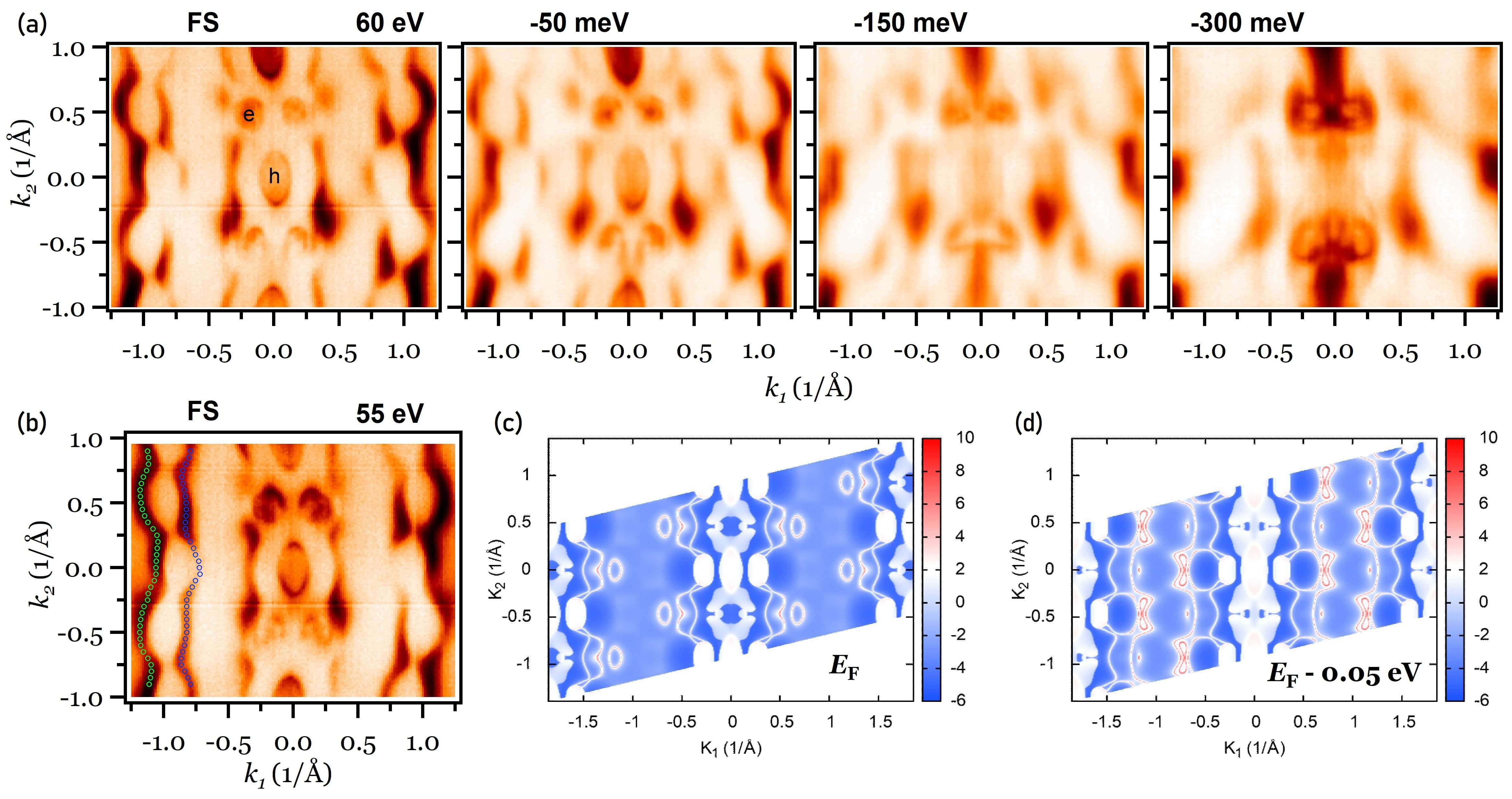}
 	\caption{ {FS and evolution of band pockets with binding energy. } (a) FS (left most panel) and constant energy contours  of $\mathrm{TaAs}_2$ measured with 60~eV photon energy. Binding energies are given on top of the plots. (b) Experimental FS measured using a photon energy of 55 eV. (c)-(d) Calculated FS  and constant energy contour at 50 meV binding energy. }
 	\label{Fig3}
 \end{figure*}

\begin{figure*}
	\centering
	\includegraphics[width=1\textwidth]{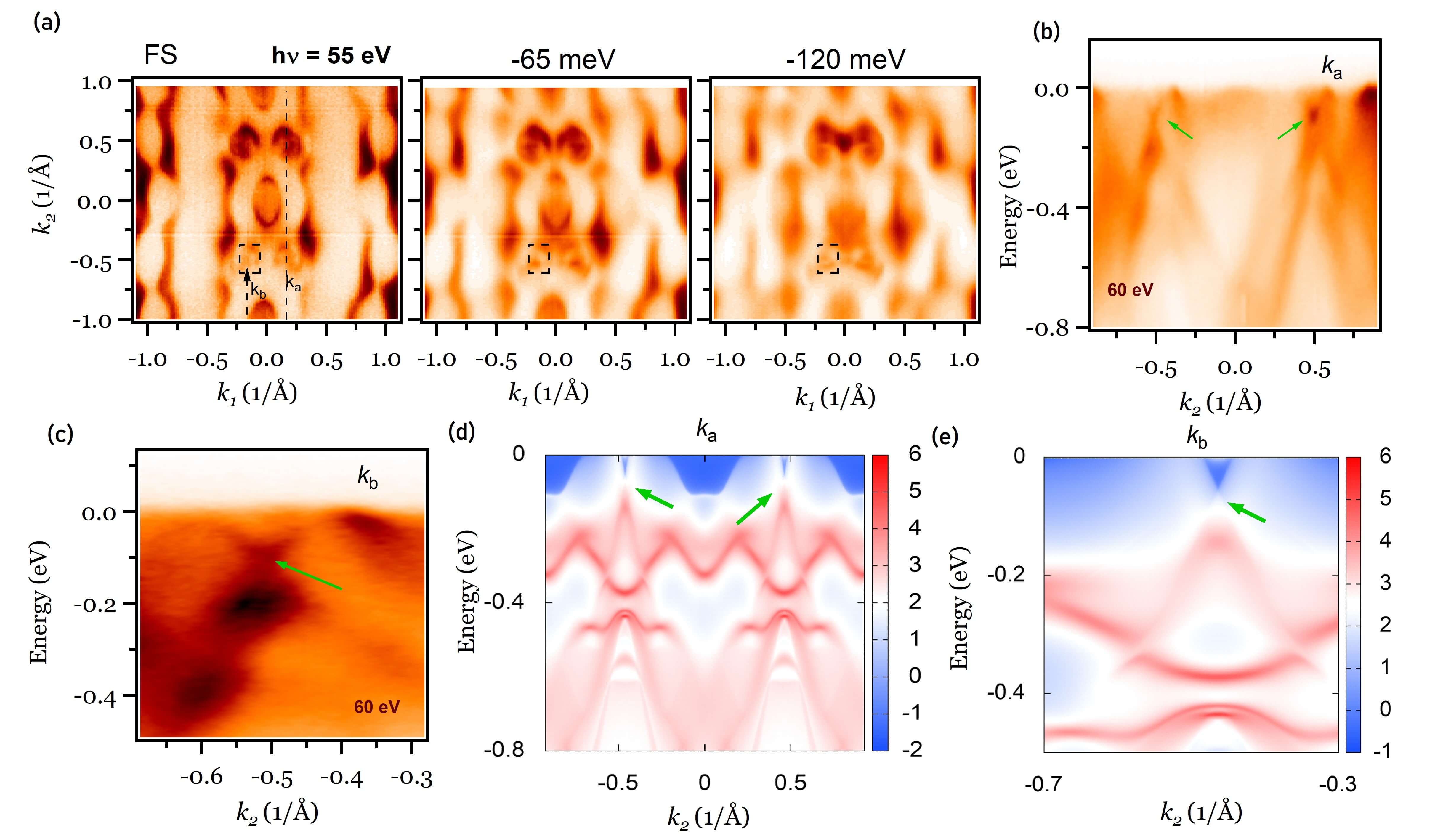}
	\caption{Observation of the linearly crossing features along the generic momentum directions. (a) FS (left) and constant energy contours (right) measured at a photon energy of 55 eV. At around 65 meV binding energy, electron pockets evolve into a point like feature (follow the black dashed box). (b)-(c) Experimental dispersion maps along k$_a$ (b) and k$_b$ (c) as indicated by dashed line in  the FS map in Fig. \ref{Fig4}(a) leftmost panel.  Observed linear crossings correspond to the momentum position of two electron like pockets, shown by green arrows. (d)-(e) Calculated bulk band structure along k$_a$ and k$_b$.}
	\label{Fig4}
\end{figure*}

Figure \ref{Fig1}(f) showcases the calculated bulk band structure without the spin-orbit coupling (SOC) effects. Contributions from various orbitals to the band structure are distinguished by using different colors. Three band crossings are identified at distinct momentum positions along the M-A and $\Gamma$-Y directions, which form nodal-lines across A and nodal-loops near M (See Ref. \cite{Wang2019} for the schematic  illustration of these nodal lines and loops). The inclusion of SOC leads to the band inversion mainly between Ta $d_{z^2}$ and Ta $d_{yz}$ orbitals at the Y and A points and the nodal lines/loops are gapped out (Fig. \ref{Fig1}(g)). The size of the SOC gap is around 300 meV (also see Ref. \cite{Wang2019}). \\

Result of Laue diffraction to identify the orientation of the cleaved surface is presented in Fig. \ref{Fig2} and results of the ARPES measurements are presented in Figs. \ref{Fig3}-\ref{Fig4} to reveal the detailed electronic structure of TaAs$_2$. The experimental Laue diffraction pattern is seen to match well with the QLaue simulation for the ($\overline{2}01$) plane, indicating that the favorable cleaving surface of $\mathrm{TaAs}_2$ is the ($\overline{2}01$) surface, which also has the lowest cleavage energy \cite{Wadge2022}. This surface is different from the (010) surface that is predicted to host rotational crystalline-symmetry-protected surface states \cite{Wang2019}. Figure \ref{Fig3}(a) shows the FS (leftmost panel) and the constant energy contours measured using a photon energy of 60 eV. The FS comprises of several features depicting a metallic nature. A similar FS is observed with measurements using a photon energy of 55 eV.  Following the band evolution while moving towards higher binding energies, we see that the elliptical (circular) pocket indicated by h (e) on the FS  increases (decreases up to 50 meV) in size confirming the hole (electron)-like nature of the associated bands near the Fermi level. Figures \ref{Fig3}(c) and \ref{Fig3}(d) show the calculated FS and the constant energy contour at a binding energy of 50 meV, respectively, which show the presence of electron and hole pockets as observed in Fig. \ref{Fig3}(a).  In the experimental FSs in Figs. \ref{Fig3}(a) and \ref{Fig3}(b), two non-closing FS features seem to extend along the $k_2$ direction within $|k_1| = \mathrm{0.75~1/\AA}$ and $|k_1|=\mathrm{1.25~1/\AA}$ (traced by green and blue circles in Fig. \ref{Fig3}(b), which represent the peak positions in the $k_1$ versus intensity plot). Open FS features have also been reported in isostructural $\mathrm{NbAs}_2$ \cite{Lou2022} and $\mathrm{MoAs}_2$ \cite{Lou2017}.\\

We now present the low photon energy dispersion maps along the directions represented by k$_a$ and k$_b$ around the electron like pockets observed in the FS (Fig. \ref{Fig4}(a)). By following the dashed black box in Fig. \ref{Fig4}(a) while moving towards higher binding energies, we can see that the small circular pocket evolves into almost a point (see -65 meV constant energy contour) and the circle reappears, as seen in the constant energy contour at around 120 meV below the Fermi level, which corresponds to the lower part of a linear crossing feature. Along the k$_a$ direction, which cuts through the two electron-like pockets, we can clearly observe two linear crossings, shown by green arrows in Fig. \ref{Fig4}(b). Figure \ref{Fig4}(c) shows the dispersion map along the k$_b$ momentum direction. Once again, we observe a similar linear crossing. The observation of the linear crossing along k$_a$ and k$_b$ nicely match with linear feature in the calculated dispersion maps (Figs. \ref{Fig4}(d) and (e)).  Comparison with the  calculated results along this generic momentum direction clearly indicates that the  bands associated with the linear crossing feature are  formed by  trivial bulk bands.\\

\begin{figure*}
	\centering
	\includegraphics[width=1\textwidth]{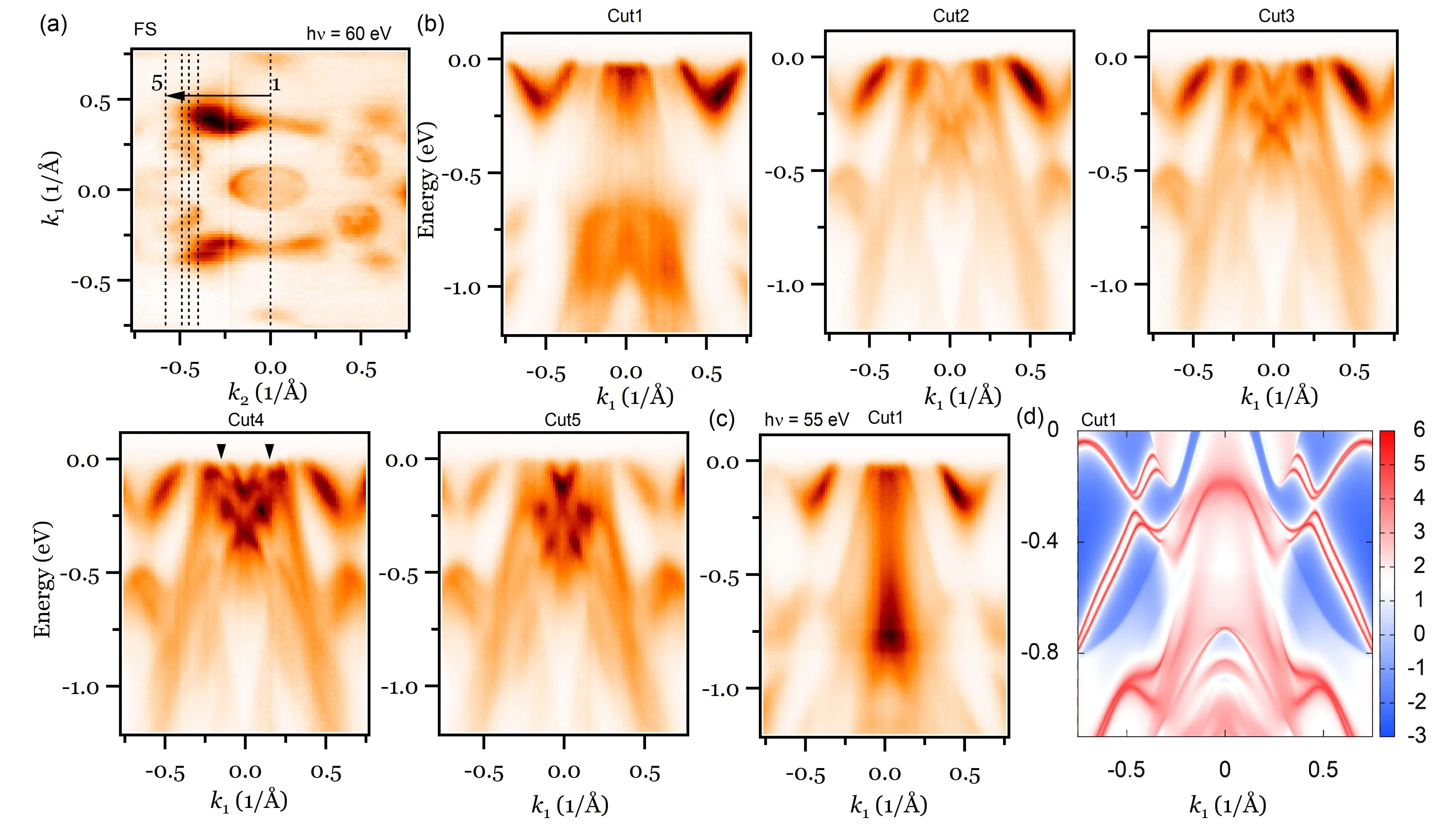}
	\caption{Dispersion maps along various directions. (a) FS map measured at 60 eV photon energy. The dashed lines show the momentum cut directions 1$\rightarrow$5  along which the dispersion maps in (b) are taken.  (c) Experimental dispersion map along cut1 measured with 55 eV photon energy.  (d) Calculated band dispersion along the cut1 direction direction.}
	\label{Fig5}
\end{figure*}

Next, we discuss the dispersion of the bands along the momentum directions indicated by the cuts 1$\rightarrow$5 in Fig. \ref{Fig5}(a). At the center of the elliptical pocket around the center of the BZ (cut1), we clearly observe hole-like bands crossing the Fermi level.  An electron band can also be seen dropping below the Fermi level away from the zone center on either side. Moving toward cut2, a  linear crossing-like feature can be seen where the lower part of this feature overlaps with another bulk band. Moving away from cut2, the linear crossing disappears and a gap appears. Along cut3 (and cut4, which lies in next BZ), the gap is at its largest and an electron-band-like feature appears with its minimum lying slightly below the Fermi level spanning between the two electron pockets (marked by black arrows for cut4 in Fig.~\ref{Fig5}(b)). Along cut5, which lies in the next BZ,  we again observe the linear-crossing feature.  The dispersion map along the cut1 direction, measured with a different photon energy of 55 eV is presented in Fig. \ref{Fig5}(c). It shows features that are similar to those seen the first panel of Fig. \ref{Fig5}(b). The calculated surface spectrum along the cut1 direction (Fig.~\ref{Fig5}(c)) is in good overall accord with the experimental results.\\

The $\mathbb{Z}_2$  topological invariants ($\nu_0$:$\nu_1\nu_2\nu_3$) in TaAs$_2$ are found to be (0;111) indicating its weak non-trivial topology \cite{Wang2019, Xu2016, Luo2016}. The (010) surface is predicted to support topological surface states protected by two-fold rotational symmetry,  giving rise to a rotational-symmetry-protected TCI phase \cite{Wang2019}. Our crystals, however, are cleaved along the ($\overline{2}01$) plane, which, unlike the (010) surface, doesnot preserve the rotational symmetry. Therefore, this surface is not expected to host surface Dirac states, which is indeed found to be the case in our experiments. This surface is thus a topologically dark surface in this weak TI material. Furthermore, the observed linear crossings do not lie at the TRIM points and correlate well with the computed bulk bands, indicating their bulk origin.  The presence of open features on the FS suggests that open-orbit fermiology could also play a role in generating extremely large MR in $\mathrm{TaAs}_2$, similar to what has been reported in other isostructural TMDs \cite{Lou2017, Lou2022}. \\

In conclusion, we report an in-depth ARPES study of the electronic structure of $\mathrm{TaAs}_2$ along with parallel first-principles calculations. The FS is found to contain multiple open and closed pockets. Several linearly crossing features near the Fermi energy are observed in the ARPES spectra, but our analysis indicates that these features are connected with bulk bands. Our study shows that the  ($\overline{2}01$) surface of $\mathrm{TaAs}_2$, which is a weak TI/TCI candidate material, is topologically dark in that it does not support crystalline symmetry protected surface states. The presence of open Fermi surface pockets suggests that the open-orbit fermiology could also contribute to the extremely large MR of $\mathrm{TaAs}_2$. \\

\noindent \textit{Acknowldgements - } M.N. is supported by the National Science Foundation CAREER award DMR-1847962 and the Air Force Office of Scientific Research MURI (FA9550-20-1-0322). The work at Northeastern University was supported by the Air Force Office of Scientific Research under award number FA9550-20-1-0322, and it benefited from the computational resources of Northeastern University's Advanced Scientific Computation Center and the Discovery Cluster. D.K. was supported by National Science Centre (NCN, Poland) under Project no. 2021/41/B/ST3/01141.  Work in the Materials Science Division of Argonne National Laboratory (sample preparation and transport characterization) was supported by the U.S. Department of Energy Office of Science, Basic Energy Sciences, Materials Sciences and Engineering Division. The work at TIFR Mumbai is supported by the Department of Atomic Energy of the Government of India. This research used resources of ALS in the Lawrence Berkeley National Laboratory, a U.S. Department of Energy Office of Science User Facility under contract number DE-AC02-05CH11231. We thank Sung-Kwan Mo for the beamline assistance at ALS.\\


\begin{thebibliography}{99}
	\bibitem{HasanKane2010}  Hasan M Z and  Kane C L 2010 \textit{Rev. Mod. Phys.} \textbf{82} 3045 \href{https://doi.org/10.1103/RevModPhys.82.3045}{10.1103/RevModPhys.82.3045}
	
	\bibitem{QiZhang2011}   Qi X-L and  Zhang S-C 2011 \textit{Rev. Mod. Phys.} \textbf{83} 1057 \href{https://doi.org/10.1103/RevModPhys.83.1057}{10.1103/RevModPhys.83.1057}
	
	
\bibitem{HasanXuNeupane2015}  Hasan M Z,  Xu S-Y and  Neupane M 2015 in \textit{Topological Insulators: Fundamentals and Perspectives}, edited by F. Ortmann, S. Roche, and S. O. Valenzuela (Wiley, Hoboken, NJ)

\bibitem{BansilLinDas2016}  Bansil A,  Lin H, and  Das T 2016  \textit{Rev. Mod. Phys.} \textbf{88} 021004  \href{https://doi.org/10.1103/RevModPhys.88.021004}{10.1103/RevModPhys.88.021004}

	\bibitem{FuKane2007}  Fu L and  Kane C L 2007 \textit{Phys. Rev. B} \textbf{76}, 045302 \href{https://doi.org/10.1103/PhysRevB.76.045302}{10.1103/PhysRevB.76.045302}
	
	\bibitem{FuKaneMele2007}  Fu L,  Kane C L, and  Mele E J 2007 \textit{Phys. Rev. Lett.} \textbf{98}, 106803 \href{https://doi.org/10.1103/PhysRevLett.98.106803}{10.1103/PhysRevLett.98.106803}
	
	\bibitem{Xia2009}  Xia Y \textit{et al} 2009 \textit{Nat. Phys.} \textbf{5}, 398 \href{https://doi.org/10.1038/nphys1274}{10.1038/nphys1274}
	
	\bibitem{Zhang2009}  Zhang H \textit{et al} 2009 \textit{Nat. Phys.} \textbf{5} 438 \href{https://doi.org/10.1038/nphys1270}{10.1038/nphys1270}
	
	\bibitem{Chen2009}  Chen Y L \textit{et al} 2009 \textit{Science} \textbf{325} 178 \href{https://doi.org/10.1126/science.1173034}{10.1126/science.1173034}
	
	\bibitem{Noguchi2019}  Noguchi R \textit{et al} 2019 \textit{Nature} \textbf{566} 518 \href{https://doi.org/10.1038/s41586-019-0927-7}{10.1038/s41586-019-0927-7}
	
	\bibitem{Fu2011}  Fu L 2011 \textit{Phys. Rev. Lett.} \textbf{106} 106802 \href{https://doi.org/10.1103/PhysRevLett.106.106802}{10.1103/PhysRevLett.106.106802}
	
	\bibitem{Hsieh2012}  Hsieh T H 2012 \textit{Nat. Commun.} \textbf{3} 982 \href{https://doi.org/10.1038/ncomms1969}{10.1038/ncomms1969}
	
	\bibitem{Tanaka2012} Tanaka  Y 2012 Nat. Phys. \textbf{8} 800 \href{https://doi.org/10.1038/nphys2442}{10.1038/nphys2442}
	
	\bibitem{Xu2012}  Xu S-Y  \textit{et al} 2012 \textit{Nat. Commun.} \textbf{3} 1192 \href{https://doi.org/10.1038/ncomms2191}{10.1038/ncomms2191}

\bibitem{Zhou2018}   Zhou X \textit{et al} 2018 \textit{Phys. Rev. B} \textbf{98} 241104(R) \href{https://doi.org/10.1103/PhysRevB.98.241104}{10.1103/PhysRevB.98.241104}

\bibitem{Hsu2019}  Hsu C-H \textit{et al} 2019 2D Mater. \textbf{6} 031004 \href{https://doi.org/10.1088/2053-1583/ab1607}{10.1088/2053-1583/ab1607}

\bibitem{Wang2019}  Wang B \textit{et al} 2019 Phys. Rev. B \textbf{100} 205118 \href{https://doi.org/10.1103/PhysRevB.100.205118}{10.1103/PhysRevB.100.205118}

\bibitem{Xu2016} Xu C \textit{et al}  2016 \textit{Phys. Rev. B} \textbf{93} 195106 \href{https://doi.org/10.1103/PhysRevB.93.195106}{10.1103/PhysRevB.93.195106}

\bibitem{Luo2016}  Luo Y \textit{et al} 2016 \textit{Sci. Rep.} \textbf{6} 27294 \href{https://doi.org/10.1038/srep27294}{10.1038/srep27294}

\bibitem{Autes2016}  Autès G \textit{et al} 2016 \textit{Phys. Rev. Lett.} \textbf{117} 066402 \href{https://doi.org/10.1103/PhysRevLett.117.066402}{10.1103/PhysRevLett.117.066402}

\bibitem{Gresch2017}  Gresch D,  Wu Q,  Winkler G W and  Soluyanov A A 2017 \textit{New J. Phys.} \textbf{19} 035001 \href{https://doi.org/10.1088/1367-2630/aa5de7}{10.1088/1367-2630/aa5de7}

\bibitem{Chen2017}  Chen J,  Li Y-K,  Dai J and C Cao 2017 \textit{Sci. Rep.} \textbf{7} 10491 \href{https://doi.org/10.1038/s41598-017-10939-1}{10.1038/s41598-017-10939-1}

\bibitem{Shao2019}  Shao Y \textit{et al} 2019 \textit{Proc. Natl. Acad. Sci.} \textbf{116} 1168 \href{https://doi.org/10.1073/pnas.1809631115}{10.1073/pnas.1809631115}

\bibitem{Sims2020}  Sims C \textit{et al} 2020 \textit{Phys. Rev. Mater.} \textbf{4} 054201 \href{https://doi.org/10.1103/PhysRevMaterials.4.054201}{10.1103/PhysRevMaterials.4.054201}

\bibitem{Bannies2021} Bannies  J \textit{et al}  2021 \textit{Phys. Rev. B} \textbf{103} 155144 \href{https://doi.org/10.1103/PhysRevB.103.155144}{10.1103/PhysRevB.103.155144}

\bibitem{Wang2014}  Wang K,  Graf D,  Li L,   Wang L and  Petrovic C 2014 \textit{Sci. Rep.} \textbf{4} 7328 \href{https://doi.org/10.1038/srep07328}{10.1038/srep07328}

\bibitem{Wang2016}  Wang Y-Y,  Yu Q-H,  Guo P-J,  Liu K, and  Xia T-L 2016 \textit{Phys. Rev. B} \textbf{94} 041103(R) \href{https://doi.org/10.1103/PhysRevB.94.041103}{10.1103/PhysRevB.94.041103}

\bibitem{Wu2016}  Wu D \textit{et al} 2016\textit{Appl. Phys. Lett.} \textbf{108} 042105 \href{https://doi.org/10.1063/1.4940924}{10.1063/1.4940924}

\bibitem{Yuan2016}  Yuan Z,  Lu H,  Liu Y,   Wang J, and  Jia S 2016 \textit{Phys. Rev. B} \textbf{93} 184405 \href{https://doi.org/10.1103/PhysRevB.93.184405}{10.1103/PhysRevB.93.184405}

\bibitem{Li2016}  Li Y \textit{et al}  2016 \textit{Phys. Rev. B} \textbf{94} 121115(R) \href{https://doi.org/10.1103/PhysRevB.94.121115}{10.1103/PhysRevB.94.121115}

\bibitem{Shen2016}  Shen B,  Deng X,  Kotliar G, and  Ni N 2016 \textit{Phys. Rev. B} \textbf{93} 195119 \href{https://doi.org/10.1103/PhysRevB.93.195119}{10.1103/PhysRevB.93.195119}

\bibitem{Kumar2017}  Kumar N \textit{et al} 2017 \textit{Nat. Commun.} \textbf{8}, 1642 \href{https://doi.org/10.1038/s41467-017-01758-z}{10.1038/s41467-017-01758-z}

\bibitem{Lou2017}  Lou R \textit{et al} 2017 \textit{Phys. Rev. B} \textbf{96}, 241106(R) \href{https://doi.org/10.1103/PhysRevB.96.241106}{10.1103/PhysRevB.96.241106}

\bibitem{Singha2018}  Singha R,  Pariari A,  Gupta G K,  Das T, and  Mandal P 2018 \textit{Phys. Rev. B} \textbf{97} 155120 \href{https://doi.org/10.1103/PhysRevB.97.155120}{10.1103/PhysRevB.97.155120} 

\bibitem{Yokoi2018}  Yokoi K \textit{et al}  2018 \textit{Phys. Rev. Mater.} \textbf{2} 024203 \href{https://doi.org/10.1103/PhysRevMaterials.2.024203}{10.1103/PhysRevMaterials.2.024203} 

\bibitem{Butcher2019}  Butcher T A \textit{et al}  2019 \textit{Phys. Rev. B} \textbf{99} 245112 \href{https://doi.org/10.1103/PhysRevB.99.245112}{10.1103/PhysRevB.99.245112} 

\bibitem{Dhakal2021}  Dhakal G \textit{et al}  2021 \textit{Phys. Rev. Research} \textbf{3} 023170 \href{https://doi.org/10.1103/PhysRevResearch.3.023170}{10.1103/PhysRevResearch.3.023170} 

\bibitem{Chen2021}  Chen S \textit{et al} 2021  \textit{Chinese Phys. Lett.} \textbf{38} 017202 \href{https://doi.org/10.1088/0256-307X/38/1/017202}{10.1088/0256-307X/38/1/017202} 

\bibitem{Ali2014}  Ali M N \textit{et al} 2014 \textit{Nature} \textbf{514} 205 \href{https://doi.org/10.1038/nature13763}{10.1038/nature13763} 

\bibitem{Zeng2016}   Zeng L K \textit{et al}  2016 \textit{Phys. Rev. Lett.} \textbf{117} 127204 \href{https://doi.org/10.1103/PhysRevLett.117.127204}{10.1103/PhysRevLett.117.127204} 

 \bibitem{Hosen2020}  Hosen M M \textit{et al} 2020 \textit{Sci. Rep.} \textbf{10} 12961 \href{https://doi.org/10.1038/s41598-020-69414-z}{10.1038/s41598-020-69414-z} 
 
 \bibitem{Wu2020}  Wu B,  Barrena V,  Suderow H, and  Guillamón I 2020 \textit{Phys. Rev. Res.} \textbf{2} 022042(R) \href{https://doi.org/10.1103/PhysRevResearch.2.022042}{10.1103/PhysRevResearch.2.022042} 
 
 \bibitem{Tafti2016}  Tafti F F,  Gibson Q D,  Kushwaha S K,  Haldolaarachchige N, and   Cava R J 2016 \textit{Nat. Phys.} \textbf{12} 272 \href{https://doi.org/10.1038/nphys3581}{10.1038/nphys3581} 
 
 \bibitem{Mun2012} Mun  E \textit{et al} 2012 \textit{Phys. Rev. B} \textbf{85} 035135 \href{https://doi.org/10.1103/PhysRevB.85.035135}{10.1103/PhysRevB.85.035135} 
 
 \bibitem{Wang2017}  Wang J \textit{et al} 2017 \textit{Sci. Rep.} \textbf{7} 15669  \href{https://doi.org/10.1038/s41598-017-15962-w}{10.1038/s41598-017-15962-w} 
 
 \bibitem{Matin2018}  Matin M,  Mondal R,  Barman N, Thamizhavel A, and  Dhar S K 2018 Phys. Rev. B \textbf{97} 205130 \href{https://doi.org/10.1103/PhysRevB.97.205130}{10.1103/PhysRevB.97.205130} 
 
  \bibitem{HK64}  Hohenberg P and  Kohn W 1964 \textit{Phys. Rev.} \textbf{136} B864 \href{https://doi.org/10.1103/PhysRev.136.B864}{10.1103/PhysRev.136.B864} 
  
  \bibitem{KS65} Kohn W and Sham L J 1965 \textit{Phys. Rev.} \textbf{140} A1133 \href{https://doi.org/10.1103/PhysRev.140.A1133}{10.1103/PhysRev.140.A1133}
  
  \bibitem{Blochl94} Blöchl P E 1994 \textit{Phys. Rev. B} \textbf{50} 17953  \href{https://doi.org/10.1103/PhysRevB.50.17953}{10.1103/PhysRevB.50.17953}
 
 \bibitem{Kresse1}  Kresse  G and  Hafner J 1993 \textit{Phys. Rev. B} \textbf{47} 558(R) \href{https://doi.org/10.1103/PhysRevB.47.558}{10.1103/PhysRevB.47.558} 
 
 \bibitem{Kresse2}   Kresse G and  Furthmüller J 1996 \textit{Comput. Mater. Sci.} \textbf{6} 15 \href{https://doi.org/10.1016/0927-0256(96)00008-0}{10.1016/0927-0256(96)00008-0} 
  
 \bibitem{Kresse3}  Kresse G and  Furthmüller J 1996 \textit{Phys. Rev. B} \textbf{54} 11169 \href{https://doi.org/10.1103/PhysRevB.54.11169}{10.1103/PhysRevB.54.11169} 
 
 \bibitem{Kresse4}   Kresse G and  Joubert D 1999 \textit{Phys. Rev. B} \textbf{59} 1758 \href{https://doi.org/10.1103/PhysRevB.59.1758}{10.1103/PhysRevB.59.1758} 
 
 \bibitem{PBE}   Perdew J P,  Burke K, and  Ernzerhof M 1996 Phys. Rev. Lett. \textbf{77} 3865 \href{https://doi.org/10.1103/PhysRevLett.77.3865}{10.1103/PhysRevLett.77.3865}
 
 \bibitem{Mostofi2008} A. A. Mostofi \textit{et al} 2008 \textit{Comput. Phys. Commun.} \textbf{178} 685 \href{https://doi.org/10.1016/j.cpc.2007.11.016}{10.1016/j.cpc.2007.11.016}
 
 \bibitem{Wu2018} Q. Wu \textit{et al} 2018 \textit{Comput. Phys. Commun} \textbf{224} 405 \href{https://doi.org/10.1016/j.cpc.2017.09.033}{10.1016/j.cpc.2017.09.033}
 
 \bibitem{YWang2016}  Wang Y \textit{et al}  2016 \textit{Nat. commun.} \textbf{7} 13142 \href{https://doi.org/10.1038/ncomms13142}{10.1038/ncomms13142} 
 
 \bibitem{Wadge2022}  Wadge A S \textit{et al}  2022 \textit{J. Phys. Condens. Matter} \textbf{34} 125601 \href{https://doi.org/10.1088/1361-648X/ac43fe}{10.1088/1361-648X/ac43fe} 
 
  \bibitem{Lou2022}  Lou R \textit{et al} 2022 \textit{Appl. Phys. Lett.} \textbf{120} 123101 \href{https://doi.org/10.1063/5.0087141}{10.1063/5.0087141}
 
\end{thebibliography}
\end{document}